\documentclass[journal=nalefd,layout=twocolumn]{achemso}
\usepackage[version=3]{mhchem} 
\newcommand*\mycommand[1]{\texttt{type=communication}}
\usepackage{amsfonts}

\author{Shuyuan Liu}
\affiliation[HYU]
{Department of Physics, Research Institute for Natural Science, and Institute for High Pressure at Hanyang
University, Hanyang University, 222 Wangsimni-ro, Seongdong-Ku, Seoul 04763, Republic of Korea}
\author{Chongze Wang}
\affiliation[HYU]
{Department of Physics, Research Institute for Natural Science, and Institute for High Pressure at Hanyang
University, Hanyang University, 222 Wangsimni-ro, Seongdong-Ku, Seoul 04763, Republic of Korea}
\author{Hyunsoo Jeon}
\affiliation[HYU]
{Department of Physics, Research Institute for Natural Science, and Institute for High Pressure at Hanyang
University, Hanyang University, 222 Wangsimni-ro, Seongdong-Ku, Seoul 04763, Republic of Korea}
\author{Jeehoon Kim}
\affiliation[POSTEC]
{Department of Physics, Pohang University of Science and Technology, Pohang 37673, Republic of Korea}
\author{Jun-Hyung Cho}
\email{chojh@hanyang.ac.kr}
\affiliation[HYU]
{Department of Physics, Research Institute for Natural Science, and Institute for High Pressure at Hanyang
University, Hanyang University, 222 Wangsimni-ro, Seongdong-Ku, Seoul 04763, Republic of Korea}

\title[\textsf{achemso}]{Interlayer Exchange Interaction Driven Topological Phase Transition in Antiferromagnetic Electride Gd$_2$O}


\begin{document}

\begin{tocentry}
\begin{center}
\includegraphics[width=8cm]{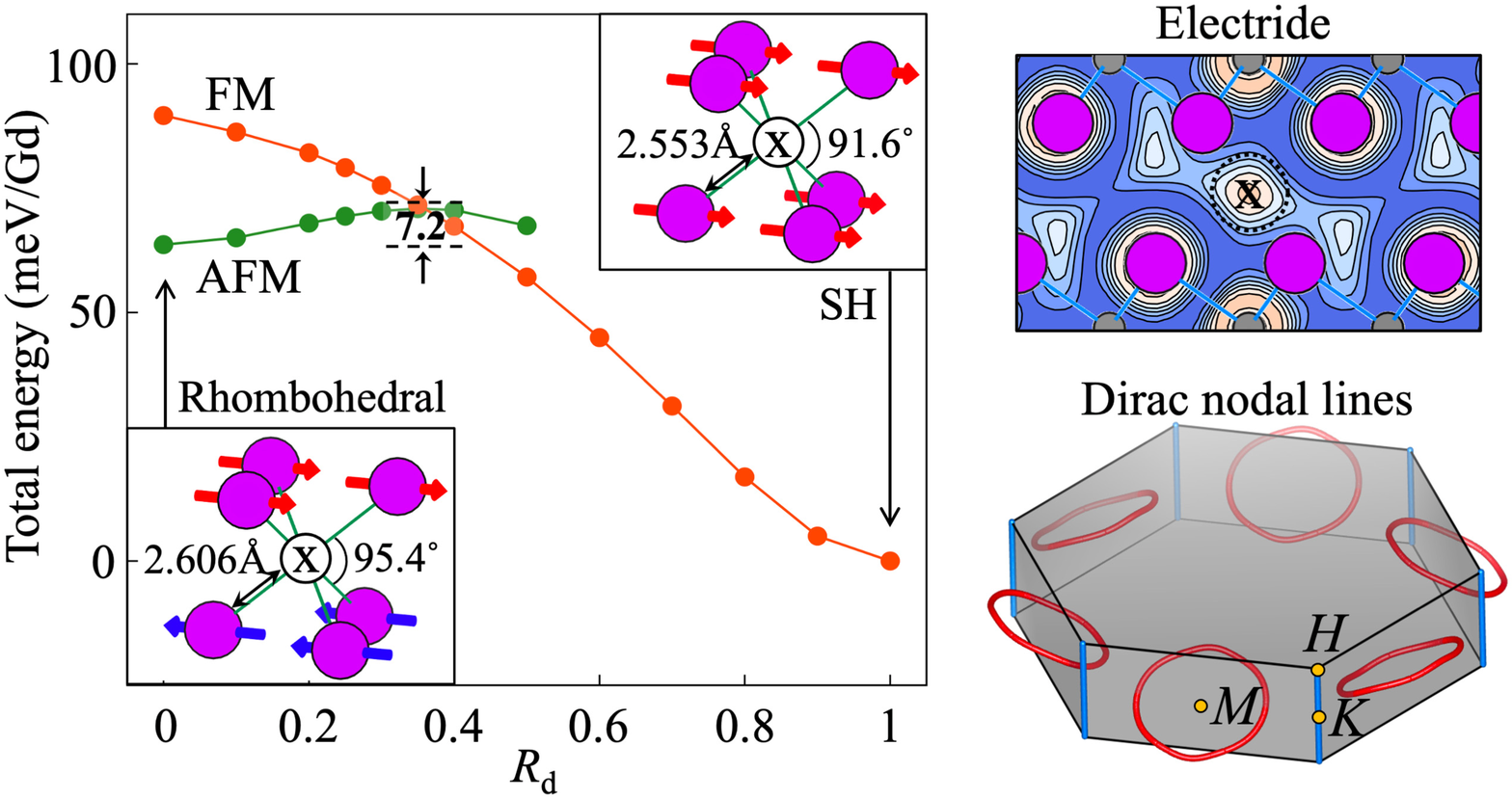}
\end{center}
\end{tocentry}

\begin{abstract}
Based on first-principles calculations, we discover a two-dimensional layered antiferromagnetic (AFM) electride Gd$_2$O, where anionic excess electrons exist in the interstitial spaces between positively charged cationic layers. It is revealed that each cationic layer composed of three-atom-thick Gd$-$O$-$Gd stacks has in-plane ferromagnetic and out-of-plane AFM superexchange interactions between the localized Gd 4$f$ spins through O 2$p$ orbitals. Furthermore, the interlayer superexchange mediated by the hybridized Gd-5$d$ and interstitial-$s$-like states involves intimate couplings between the spin, lattice, and charge degrees of freedom, thereby inducing simultaneous magnetic, structural, and electronic phase transitions. The resulting ground state with the simple hexagonal lattice hosts massless Dirac fermions protected by nonsymmorphic magnetic symmetry, as well as massive Dirac fermions. We thus demonstrate that the anionic excess electrons in Gd$_2$O play a crucial role in the emergence of magnetic Dirac semimetal states, therefore offering an intriguing interplay between 2D magnetic electrides and topological physics.
\end{abstract}
\vspace{0.5cm}

Electrides are a special kind of ionic crystals, where excess electrons residing in interstitial regions behave as anions~\cite{Dye-Science1990,Electride-Nature1993,Dye-Science2003,review2021-1,review2021-2}. Such anionic excess electrons interact with cationic frameworks, attaining the structural stability of electrides~\cite{Electride-Hoffman-JACS2015}. Historically, earlier crystalline electrides stemmed from organic materials~\cite{organicElectride-JACS1983}. Since the organic electrides tend to be stabilized only at low temperatures with an inert gas atmosphere, it has been hampered to utilize their intrinsic electronic properties for practical applications. By contrast, inorganic electrides not only have high thermal and chemical stability, but also provide exciting properties, such as high electrical conductivities, low work functions, highly anisotropic optical response, and rich surface-related chemical reactions~\cite{C12A7-Science2003,catalysis-Nat.Chem2012,Ca2N-Nature2013}. The first inorganic electride [Ca$_{24}$Al$_{28}$O$_{64}$]$^{4+}$4$e^-$ synthesized in 2003~\cite{C12A7-Science2003} exhibited various electronic phases ranging from insulator to superconductor, because excess electrons trapped in the clathrate Ca$-$Al$-$O cages gradually delocalize with increasing their concentrations~\cite{C12A7-SC-JACS2007,C12A7-SC-PTRSA2015}. Almost a decade ago, the layer structured inorganic electride [Ca$_2$N]$^{+}$$e^-$ consisting of a three-atom-thick building block of Ca$-$N$-$Ca stacks was synthesized~\cite{Ca2N-Nature2013}, where anionic excess electrons are confined in the interstitial spaces between positively charged [Ca$_2$N]$^{+}$ cationic layers. This new class of electride has triggered many theoretical~\cite{Electride-PRX2014,Seho,Topological-Electride-PRX2018,LiangLiu,shuyuan-jpcc} and experimental~\cite{Electride-Y2C2014,Ca2NML2016,Gd2C-Nat.Commun.2020,Hf2S-Sci. Adv.2020} efforts to develop a variety of two-dimensional (2D) layered inorganic electrides.

From the theoretical side, high-throughput computational methods have been employed to search for inorganic electrides~\cite{HP-CM2018,HP-CM2019,HP-matter2019}. In particular, the combined features of superior electride properties and magnetism are promising for various applications in spintronics, such as spin injection, long spin relaxation, and magnetic tunnel junctions. However, in contrast to the discovery of many nonmagnetic (NM) electrides~\cite{Electride-JACS2016,HP-CM2018,HP-CM2019,HP-matter2019}, magnetic electrides have been rarely discovered so far~\cite{Electride-PRX2014,magnetic-electride2021}. To the best of our knowledge, the experimentally synthesized 2D magnetic electrides are only rare earth compounds such as Y$_2$C~\cite{Electride-Y2C2014,2016Y2C,Y2CFM-JACS2017,2018Y2C} and Gd$_2$C~\cite{Gd2C-Nat.Commun.2020} that are isostructural to Ca$_2$N. Compared to a weak ferromagnetic (FM) or paramagnetic order of Y$_2$C, Gd$_2$C was observed to exhibit a strong FM order with a Curie temperature of ${\sim}$350 K~\cite{Gd2C-Nat.Commun.2020}. Subsequently, a first-principles calculation of Gd$_2$C predicted the existence of Weyl semimetal states~\cite{Gd2C-PRL2020}. However, there has been no theoretical or experimental report yet of 2D antiferromagnetic (AFM) electrides possessing topological states, which could embody faster spin dynamics and robustness stability beneficial for spintronics applications (see~\cite{AFM-topo1,AFM-topo2,AFM-topo3} for the recent reviews of AFM materials).

Recently, 2D van der Waals (vdW) magnets~\cite{2DvdW-Nature2017,2DvdW-Nature2018,2DvdW-AM2019} have been discovered to exhibit the presence of intrinsic FM and AFM ground states at finite temperatures down to atomic-layer thicknesses, thereby opening a new horizon in magnetic materials science. In such 2D magnetic systems, magnetic anisotropy is essential for a long-range magnetic order that prevails against thermal fluctuations imposed by the Mermin-Wagner theorem~\cite{Mermin-WagnerPRL1966}. Unlike 2D vdW magnets with weak interlayer interactions, 2D magnetic electrides can have strong interlayer interactions such as Ruderman–Kittel–Kasuya–Yosida~\cite{RKKY1957} or superexchange~\cite{superexchange1950} through anionic excess electrons~\cite{Gd2C-Nat.Commun.2020}. Therefore, such enhanced interlayer interactions between neighboring cationic layers may induce intimate couplings between the spin, lattice, and charge degrees of freedom in 2D magnetic electrides, as will be demonstrated below.

In this Letter, using first-principles density-functional theory (DFT) calculations, we predict that Gd$_2$O having the layered Ca$_2$N-type structure is a new 2D AFM electride, where each monolayer (ML) has in-plane FM and out-of-plane AFM superexchange interactions (see Figure 1a) through O 2$p$ orbitals. It is also found that the AFM and FM interlayer superexchange interactions mediated by the hybridized Gd-5$d$ and interstitial-$s$-like states stabilize the rhombohedral and simple hexagonal (SH) stacked structures (see Figures 1b and c), respectively. Here, the structural transition from the rhombohedral to SH phase is accompanied by an insulator-to-metal transition. Furthermore, the SH ground state hosts the coexistence of massless and massive Dirac semimetal states. It is thus demonstrated that the 2D AFM electride Gd$_2$O possesses an intricate interplay between magnetism, geometry, electricity, and topology, thereby enriching our understanding of a new class of topological magnetic electrides. The present findings are rather generic and hence, they can also be applicable to other isostructural rare-earth electrides Tb$_2$O and Dy$_2$O.

\begin{figure}[h!t]
\includegraphics[width=8.5cm]{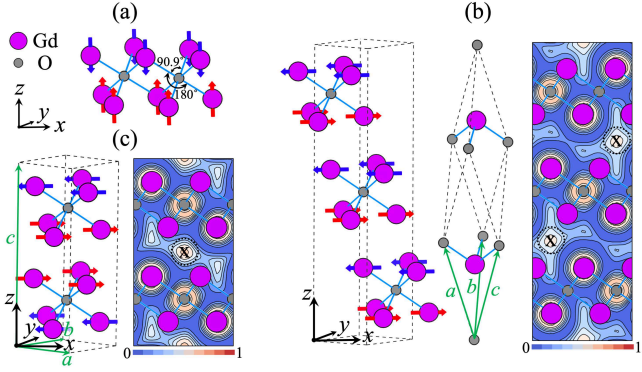}
\caption{Optimized structures of (a) ML Gd$_2$O and the (b) rhombohedral and (c) SH phases of bulk Gd$_2$O. The calculated ELFs for the rhombohedral and SH phases are also given on the (110) plane of the conventional unit cell. In (b), the conventional unit cell (left) and the primitive cell (middle) are drawn. The lattice parameters of the rhombohedral phase are $a$ = $b$ = $c$ = 6.451 {\AA}, while those of the SH phase are $a$ = $b$ = 3.560 {\AA} and $c$ = 11.641 {\AA}. The arrows represent the magnetic moments pointing parallel to out-of-plane and in-plane directions.}
\label{figure:1}
\end{figure}

We first explore the ground state of ML Gd$_2$O using the DFT + $U$ calculation without including spin-orbit coupling (SOC). The structure of ML Gd$_2$O is composed of three atomic layers, where O atoms locating in a triangular lattice are surrounded by six Gd atoms in an octahedral geometry (see Figure 1a). We find that the AFM phase having the in-plane FM and out-of-plane AFM orderings of Gd spins is more stable than the NM and FM phases by 16.115 eV and 57.5 meV per Gd atom, respectively. In this AFM phase, the calculated magnetic moment arising from the Gd 4$f$ orbitals is 6.945 ${\mu}_B$ per Gd atom, close to that of Gd$^{3+}$ ion having seven fully aligned half-filled 4$f$ electrons with the total spin of $S$ = 7/2. This indicates that the ionic magnetic moment of Gd 4$f$ orbitals is nearly intact because of their negligible spatial overlap between adjacent Gd atoms. The resulting large Gd 4$f$ spin state polarizes Gd 5$d$ electrons via the intra-atomic (Hund’s rule) exchange, which contributes an additional magnetic moment of 0.453 ${\mu}_B$ per Gd atom.

It is worth noting that the AFM ground state exhibits a good thermodynamic stability with a large formation energy of 2.028 eV per Gd atom and the absence of any imaginary phonon frequency (see Figure S1). To examine the thermodynamical stability of Gd$_2$O, we investigate its formation energy with those of other Gd oxides such as Gd$_2$O$_3$, GdO, GdO$_2$, and GdO$_3$ (see Materials Project~\cite{MP} and Inorganic Crystal Structure Database~\cite{ICSD}) using the convex hull analysis. As shown in Figure S2, we find that the degree of thermodynamic stability of Gd$_2$O is comparable to those of metastable Gd oxides that were experimentally synthesized. Furthermore, we ensure the thermodynamic stability of Gd$_2$O by using ab initio molecular dynamics simulations and find that Gd$_2$O preserves its layered structure up to ${\sim}$1000 K without any structural transformation (see Figure S3).

Figure 2a shows the AFM band structure of ML Gd$_2$O, where the Gd 4$f$ and O 2$p$ states are located around $-$8.5 and $-$5.7 eV below $E_F$, respectively. The calculated partial density of states (PDOS) shows some hybridization of the O 2$p$ states with the Gd 5$d$ states (see Figure 2a). According to the Goodenough-Kanamori-Anderson (GKA) rules~\cite{superexchange1950,GKA-Goodenough,GKA-Kanamori} for superexchange interactions, the Gd 4$f$ spins between the top and bottom layers with the Gd$-$O$-$Gd bond angle of 180$^{\circ}$ are antiferromagnetically coupled via the $p-d$ orbital hybridization, while on either the top or bottom layer, the Gd 4$f$ spins with the Gd$-$O$-$Gd bond angle of 90.9$^{\circ}$ favor a FM coupling (see Figure 1a). The resulting in-plane FM and out-of-plane AFM configuration breaking inversion symmetry leads to the opening of a slightly negative gap, as shown in the inset of Figure 2a. However, this negative gap is converted into the positive one by using the hybrid calculation (see Figure S4).

\begin{figure}[h!t]
\includegraphics[width=8.5cm]{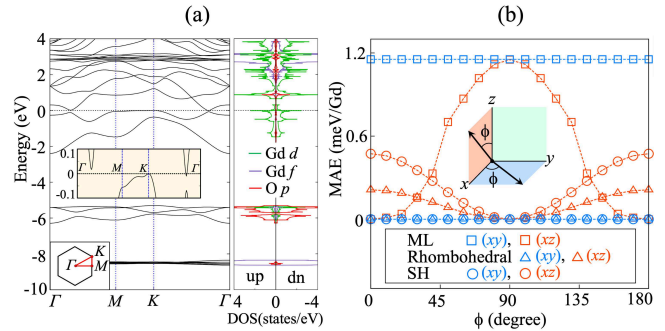}
\caption{(a) Calculated band structure of ML Gd$_2$O, together with the PDOS for the Gd $d$, $f$, and O $p$ orbitals. The insets in (a) show the close-up band structure near $E_F$ and the 2D Brillouin zone (BZ) of ML Gd$_2$O. In (b), the angular dependences of MAE with respect to ${\phi}$ on the $xy$- and $xz$-planes are displayed for ML as well as the rhombohedral and SH bulk phases. Here, the energy of each ground state is set to zero.}
\label{figure:2}
\end{figure}

By including SOC, we calculate the magnetic anisotropy energy (MAE) of ML Gd$_2$O. Figure 2b displays the angular dependence of MAE on the $xy$- and $xz$-planes (see Figure S5 for the result on the $yz$-plane). We find that the MAE on the $xy$-plane is isotropic, whereas that on the $xz$- or $yz$-plane strongly depends on the angle ${\phi}$ relative to the $z$ direction. As a result, the easy axis points along the $z$ direction (see Figure 1a) with a MAE of 1.152 meV per Gd atom. Interestingly, this out-of-plane easy-axis direction is changed into in-plane directions in bulk Gd$_2$O, implying that the interlayer interactions via anionic excess electrons drastically influence magnetic anisotropy, as discussed below.

Next, we study the electronic and magnetic properties of bulk Gd$_2$O having the rhombohedral structure (see Figure 1b) which is the typical crystal structure of other 2D layered electrides such as Ca$_2$N~\cite{Ca2N-Nature2013}, Y$_2$C~\cite{Electride-Y2C2014}, and Gd$_2$C~\cite{Gd2C-Nat.Commun.2020,Gd2C-PRL2020}. This rhombohedral phase is found to favor the AFM interlayer coupling (see Figure 1b) over the FM one. Figure 3a shows the band structure of the rhombohedral phase, obtained using DFT + $U$ with including SOC. Note that the AFM rhombohedral phase with the magnetic space group $P$-1$'$ (No. 2.6) breaks each of inversion symmetry $P$ and time-reversal symmetry $T$, but respects the combinatory symmetry $PT$. Since the square of $PT$ operator equals $-$1, each band has two-fold degenerate energy levels (i.e., the Kramers degeneracy). In order to reveal the topological feature of the rhombohedral phase, we calculate the evolution of the Wannier charge centers of the occupied valence bands (see Figure S6) using the Wilson loop approach~\cite{Wilsonloop,WCC,wannnier90,wanniertools}, and obtain a zero $Z_2$ invariant. Therefore, we can say that the rhombohedral phase is a trivial insulator (see Figure 3a).

\begin{figure}[h!t]
\includegraphics[width=8.5cm]{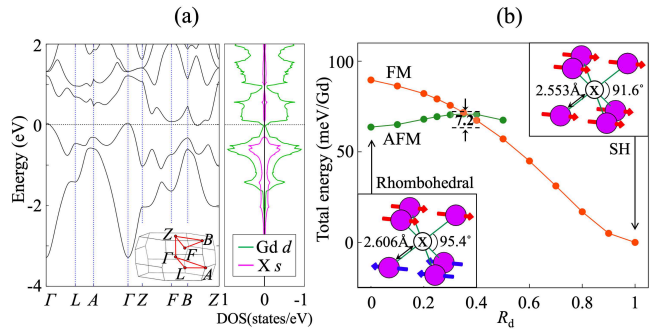}
\caption{(a) Calculated band structure of the rhombohedral phase, together with the PDOS for the Gd $d$ and interstitial-$s$-like orbitals. The inset in (a) shows the BZ of the rhombohedral structure. In (b), the energy variations of the FM and AFM rhombohedral phases are displayed as a function of lateral shear distortion $R_d$, defined as a ratio $d$/$d_0$ where $d$ represents the lateral shifted displacement from the AFM rhombohedral toward the FM SH structure. The insets in (b) show $d_{\rm Gd-X}$ and ${\theta}_{Gd-X-Gd}$ values of the rhombohedral and SH ground states around X anions. The total energy of the FM SH phase is set to zero.}
\label{figure:3}
\end{figure}

To characterize the 2D electride nature of Gd$_2$O, we calculate the electron localization function (ELF)~\cite{ELF} of the rhombohedral phase, showing that interstitial electrons are highly localized at the positions marked as ``X" in the interlayer space [see Figure 1b]. As shown in Figure 3a, such interstitial electron states exhibit a strong hybridization with the Gd 5$d$ states around $-$0.7 eV below $E_F$. Therefore, similar to the case of ML Gd$_2$O, the Gd 4$f$ spins between neighboring layers are antiferromagnetically coupled via superexchange, mediated by the hybridized Gd-5$d$ and interstitial-$s$-like states.

In contrast to the insulating nature of the AFM rhombohedral phase, the FM rhombohedral phase is metallic (see Figure S7). It is noted that the FM rhombohedral phase is dynamically unstable with the presence of imaginary phonon frequencies  (see Figure S8). As the structure of the FM rhombohedral phase is allowed to relax along the calculated forces, it converges to the SH phase (see Figure 1c) that is thermodynamically more stable than the AFM rhombohedral phase by 63.6 meV per Gd atom. Figure 3b shows the energy variations of the FM and AFM rhombohedral phases as a function of $R_d$. We find that the FM and AFM rhombohedral phases exhibit an energy crossover around $R_d$ = 0.36. However, it cautions that the phase transition from the AFM to the FM rhombohedral phase cannot occur by a thermal excitation over the energy barrier of ${\sim}$7.2 meV (see Figure 3b) because of their different spin configurations. Instead, spin flips in the AFM rhombohedral phase can occur above the N$\acute{\rm e}$el temperature~\cite{Neel}. Considering that the SH phase exhibits a FM metallic character with a strong hybridization between the Gd-5$d$ and interstitial-$s$-like states (see Figure 4a), the structural phase transition from rhombohedral to SH is accompanied by an insulator-to-metal transition. It is noted that the comparison of the projected band structures and ELFs between the SH Gd$_2$O and the rhombohedral Gd$_2$C obviously exhibits their different features in the electronic structure around $E_F$ and the distribution of interstitial anionic electrons (see Figures S9 and S10).

The insets in Figure 3b show the bond length $d_{\rm Gd-X}$ and bond angle ${\theta}_{Gd-X-Gd}$ in the rhombohedral and SH ground states. We obtain $d_{\rm Gd-X}$ = 2.606 {\AA} and ${\theta}_{Gd-X-Gd}$ = 95.4$^{\circ}$ for the rhombohedral phase, while $d_{\rm Gd-X}$ = 2.553 {\AA} and ${\theta}_{Gd-X-Gd}$ = 91.6$^{\circ}$ for the SH phase. Since ${\theta}_{Gd-X-Gd}$ of SH is close to 90$^{\circ}$, the Gd 4$f$ spins between neighboring layers favor the FM superexchange interaction in terms of the GKA rules~\cite{superexchange1950,GKA-Goodenough,GKA-Kanamori}. Meanwhile, the longer $d_{\rm Gd-X}$ value in rhombohedral likely attains a larger gain of exchange kinetic energy, thereby favoring the AFM superexchange interaction.

In Figure 2b, we find that the easy axes of the rhombohedral and SH phases lie in-plane with the MAE values of 0.216 and 0.473 meV per Gd atom, respectively. In order to understand the different orientations of easy axis between bulk and ML Gd$_2$O, we calculate their band structures and DOS with respect to ${\phi}$ relative to the $z$ direction. For the two bulk phases, the energies of the occupied bands near $E_F$ lower gradually with increasing ${\phi}$, while for ML Gd$_2$O, the energy lowering is maximized at ${\phi}$ = 0 (see Figure S11). According to the force theorem~\cite{force-theorem1,force-theorem2}, the summation of the band energies up to $E_F$ contributes to determine MAE. It is thus likely that the different variations of ML and bulk band energies with respect to ${\phi}$ are associated with their switching of easy axis. We further notice that for the bulk phases, the hybridized Gd-5$d$ and interstitial-$s$-like states enable the interatomic hopping along the $z$ axis to generate orbital magnetic moment, which may in turn contribute to the in-plane orientation of magnetization via SOC.

The SH phase having the FM interlayer coupling doubles the primitive cell (see Figure 1c) compared to the rhombohedral phase. We find that the band structure of the SH phase (see Figure 4a) exhibits the Kramers degeneracy enforced by $PT$ symmetry. Figures 4a and b show that there are two types of Dirac nodal lines (DNLs). One is the circular-shaped DNL$_c$ protected by $P$ symmetry (see also Figure S12) and the other is the linear DNL$_l$ around the $K-H$ line, protected by the threefold rotation symmetry $C_{3z}$. However, by taking into account SOC, the fourfold degeneracy of these DNLs is lifted to open gaps smaller than ${\sim}$68 meV (see Figure S13), except at the $H$ and $H'$ points where the Dirac points are protected by the nonsymmorphic symmetry $T{\tau}_{1/2}$ containing a half translation ${\tau}_{1/2}$ along the $z$ direction (see symmetry analysis in the Supporting Information~\cite{Natphy}). It is thus remarkable that the structural transformation from rhombohedral to SH gives rise to the topological transition from a trivial insulator to a Dirac semimetal having massless and massive Dirac fermions. Figure 4c shows the projected surface spectrum for the (001) surface of Gd$_2$O, obtained by using the Green’s function method based on the tight-binding Hamiltonian with maximally localized Wannier functions~\cite{wannnier90,wanniertools}. We find the existence of a spin polarized Fermi arc surface state connecting the two Dirac points (see Figure 4c). In Figure 4d, the projected Fermi surface of the (001) surface also exhibits the spin-polarized topological surface states around the ${\overline{M}}$ point with helical spin textures.

\begin{figure}[h!t]
\includegraphics[width=8.5cm]{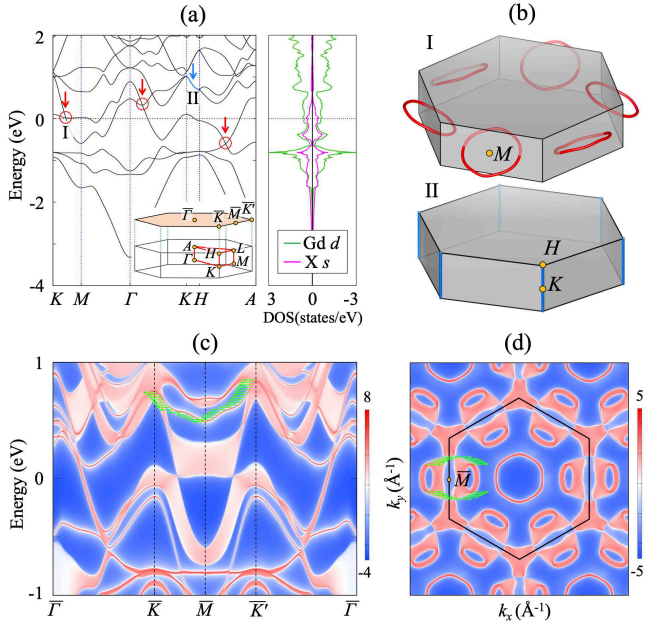}
\caption{(a) Calculated band structure and PDOS of the SH phase and (b) DNL$_c$ (DNL$_l$) residing in the region I (II). In (a), the red and blue arrows represent the crossing points in DNL$_c$ and DNL$_l$, respectively. The inset in (a) shows the BZ of the SH structure together with the projected surface BZ of the (001) surface. The projected spectrum and Fermi surface of the (001) surface are displayed in (c) and (d), respectively. In (c) and (d), the arrows represent the spin textures of surface states with only the $S_x$ (horizontal) and $S_y$ (vertical) components.}
\label{figure:4}
\end{figure}

\section{Conclusion}
Based on first-principles DFT calculations, we discover for the first time the AFM electride Gd$_2$O that has the topological Dirac states protected by nonsymmorphic magnetic symmetry, thereby providing fundamental new insights for more nanospintronics applications rooted in its unique AFM order. Specifically, this AFM electride exhibits a novel superexchange between the localized Gd 4$f$ spins via intermediate O 2$p$ orbitals as well as interstitial anion electrons, giving rise to complex interlayer magnetic couplings. Furthermore, highly mobile anionic electrons in the interstitial spaces between positively charged cationic layers play an important role in inducing intricate magnetic, structural, and electronic phase transitions, which enrich our understanding of a new class of topological AFM electrides. In addition, other isostructural lanthanide oxides such as Tb$_2$O and Dy$_2$O can also exhibit similar structural, electronic, magnetic, and topological properties as predicted in Gd$_2$O (see Figure S14). Therefore, our findings provide a venue to investigate the intriguing interplay between AFM electrides and topological physics, which will be promising for future spintronics.

\section{Methods}
The present first-principles DFT calculations were performed using the Vienna ab initio simulation package with the projector-augmented wave method~\cite{vasp1,vasp2,paw}. The exchange-correlation energy was treated with the generalized-gradient approximation functional of Perdew-Burke-Ernzerhof~\cite{pbe}. For the DFT + $U$ calculation, we used the Dudarev’s approach~\cite{Dudarev} with the Hubbard parameter $U$ = 6.7 eV and exchange interaction parameter $J$ = 0.7 eV. The plane wave basis was employed with a kinetic energy cutoff of 550 eV, and the $k$-space integration was done with 18${\times}$18${\times}$1, 12${\times}$12${\times}$12, and 18${\times}$18${\times}$6 meshes for ML, rhombohedral, and SH phases, respectively. All atoms were allowed to relax along the calculated forces until all the residual force components were less than 0.005 eV/{\AA}. The phonon spectrum calculations were carried out by using the QUANTUM ESPRESSO package~\cite{qe}, with the 6${\times}$6${\times}$1, 4${\times}$4${\times}$4, and 6${\times}$6${\times}$2 $q$ points for ML, rhombohedral, and SH phases, respectively. We note that for ML Gd$_2$O, the variation of the effective $U$ value, $U_{\rm eff}$ = $U$ $-$ $J$, significantly changes the positions of Gd 4$f$ states, but the energy dispersion of the electronic states around $E_F$ changes little with respect to $U$ (see Figure S4). Specifically, the positions of Gd 4$f$ states and the gap size at $E_F$ obtained using the DFT + $U$ calculation with $U_{\rm eff}$ = 6 eV agree well with those obtained using the hybrid calculation with the HSE functional~\cite{HSE1,HSE2} (see Figure S4).

\section{Author Information}
\textbf{Corresponding Author}\\
\textbf{Jun-Hyung Cho} $-$ Department of Physics, Research Institute for Natural Science, and Institute for High Pressure at Hanyang
University, Hanyang University, 222 Wangsimni-ro, Seongdong-Ku, Seoul 04763, Republic of Korea; Email: chojh@hanyang.ac.kr\\

\noindent \textbf{Authors}\\
\textbf{Shuyuan Liu} $-$ Department of Physics, Research Institute for Natural Science, and Institute for High Pressure at Hanyang
University, Hanyang University, 222 Wangsimni-ro, Seongdong-Ku, Seoul 04763, Republic of Korea\\
\textbf{Chongze Wang} $-$ Department of Physics, Research Institute for Natural Science, and Institute for High Pressure at Hanyang
University, Hanyang University, 222 Wangsimni-ro, Seongdong-Ku, Seoul 04763, Republic of Korea\\
\textbf{Hyunsoo Jeon} $-$ Department of Physics, Research Institute for Natural Science, and Institute for High Pressure at Hanyang
University, Hanyang University, 222 Wangsimni-ro, Seongdong-Ku, Seoul 04763, Republic of Korea\\
\textbf{Jeehoon Kim} $-$ Department of Physics, Pohang University of Science and Technology, Pohang 37673, Republic of Korea\\

\noindent \textbf{Author Contributions}\\
S. L. and C. W. contributed equally to this work.\\

\noindent \textbf{Notes}\\
The authors declare no competing financial interest.\\

\begin{suppinfo}

The Supporting Information is available free of charge at DOI: 10.1021/xxxxxx.

See Supporting Information for the symmetry and topology analyses, the calculated phonon spectra and formation energies of ML and bulk Gd$_2$O, the convex hull analysis of various Gd oxide compounds, the ab initio molecular dynamics simulations, the DFT + $U$ and HSE band structures, the angular dependence of MAE on the $yz$ plane, the $Z_2$ number for the rhombohedral phase, the projected band structures and ELFs of Gd$_2$O and Gd$_2$C, the band structure changes with respect to ${\phi}$, the SOC-induced gaps, and the band structures of Tb$_2$O and Dy$_2$O.
\end{suppinfo}

\begin{acknowledgement}

We are grateful for discussions with Dr. Hyun-Jung Kim. This work was supported by the National Research Foundation of Korea (NRF) grant funded by the Korean Government (Grants No. 2019R1A2C1002975, No. 2016K1A4A3914691, and No. 2015M3D1A1070609). The calculations were performed by the KISTI Supercomputing Center through the Strategic Support Program (Program No. KSC-2020-CRE-0163) for the supercomputing application research.

\end{acknowledgement}


\end{document}